# Modulating Endothermic Singlet Fission by Controlling Radiative Rates in Perylene Dimers

Nadezhda V. Korovina,[a] Shea O'Sullivan,[a] Jennica Kelm,[a] Yunhui L. Lin,[a] Katherine Lloyd,[b] and Justin C. Johnson[a]

[a]  Nadezhda V. Korovina, Yunhui L. Lin, Shea O'Sullivan, Jennica Kelm, Katherine Lloyd, and Justin C. Johnson
     Chemistry and Nanoscience Science Center
     National Renewable Energy Laboratory
     Golden, CO 80401 USA
     E-mail: justin.johnson@nrel.gov

[b]  Katherine Lloyd
     Department of Chemistry
     Rutgers University
     73 Warren Street, Newark, NJ 07102, USA
     Newark, NJ 07102 USA

**Abstract:** Endothermic singlet fission (SF), an exciton multiplication process that produces a pair of high-energy triplet excitons $(T_1T_1)$, is appealing for photovoltaic or photoelectrochemical applications, as it allows the conversion of entropy into electronic or chemical energy. The mechanistic aspects of this process are not entirely known, and strategies for improving the yield of triplets via endothermic SF have not been developed. In this work we provide experimental evidence that in photoexcited dimers of perylene, $S_1$ is initially in equilibrium with $^1(T_1T_1)$, and that the lifetime of this equilibrium can be controlled through strategic changes in the radiative rate. Through careful molecular design we fine-tune both the degree of endothermicity and excited state lifetimes in four perylene dimers. Using transient absorption and time-resolved fluorescence, we reveal that the dimer with the slowest radiative rate constant produces the most prolonged $^1(T_1T_1)$. However, in the dimers, the annihilation of the $^1(T_1T_1)$ state results in a single long-lived triplet rather than a pair, and increasing the free triplet yield above 100% would require additional chromophores.

## Introduction

Singlet fission (SF) is a process by which a photoexcited chromophore in the $S_1$ state shares its energy with a neighboring chromophore, resulting in a pair of triplet $(T_1)$ excitons.[1] This triplet pair is initially correlated into an overall singlet – $^1(T_1T_1)$ – and is often designed to lie energetically close to or below the $S_1$ state. As a result, SF proceeds very rapidly (ps or faster) in many molecular crystals,[2] and covalent dimers,[3] and oligomers.[4] If the $T_1$ excitons could be independently harvested in a photovoltaic device, the efficiency of such a device could exceed the Shockley-Queisser limit.[5] SF materials can obviate the typical thermalization losses of conventional PV materials by converting high energy excitons into pairs of lower energy excitons. As with any such scheme, the most energy-efficient scenario ostensibly is one where the reactant and product, which in the case of SF are $S_1$ and $^1(T_1T_1)$, are nearly isoenergetic. There, a small driving force can push reactants toward products while minimizing reversibility and without wasting potential energy. However, construction of such a scheme based solely on optical and/or electrochemical analysis of energies is ignorant of the total Gibbs energy, including both enthalpy and entropy, which is an intrinsic component of SF because the available product microstates are far more numerous than those of the reactant.[6] Therefore, an even more favorable approach is to engender endothermic SF, where the entropy harnessed during SF allows the production of $^1(T_1T_1)$ of higher energy than the photon used in photoexcitation.[7]

The most commonly studied material that undergoes weakly endothermic SF is crystalline tetracene.[8] There, the endothermicity of SF is ~50-150 meV, and the $T_1$ excitons were observed to appear in 40-90 ps.[9] The prevailing theory that explains such rapid endothermic SF in crystalline tetracene and its derivatives is that the inherent entropy associated with the SF process compensates for the electronic energy gap between $S_1$ and $^1(T_1T_1)$, allowing the overall process to be thermodynamically favorable.[6a] Of key importance to this hypothesis is the notion that spatial separation of the resulting triplet pair within the delocalization radius is the primary step in producing long-lived triplet excitons, a mechanism that has been studied in rubrene crystals[10] as well as tetracene[11] and perylene oligomers.[4b]

In our prior study of linear perylene oligomers we explored endothermic SF (~0.3 eV endothermicity) as a function of oligomer length, and we observed that pairs of long-lived triplet excitons form in solutions of the trimer and tetramer, but not the dimer. The long-lived triplet yields, however, were limited to 30% in the best-case scenario. While these perylene systems present many advantages, such as superior photostability and large extinction coefficients in the visible regions, to be useful for photovoltaic applications they must be able to produce >100 % $T_1$ upon SF. Improving upon the reported SF yields by tuning excited state parameters through molecular structure is the primary focus of this report.

An earlier study by Gilligan *et al.* explained the excited state dynamics of a rigid tetracene dimer in terms of an equilibrium between the $S_1$ and the $^1(T_1T_1)$ state populations,[12] and this is common to tetracene chromophore assemblies.[13] Although oligomers were largely the focus of our prior investigations, we have evidence to suggest an excited state equilibrium may also be at work in our perylene dimers. Lacking fast channels toward $^1(T_1T_1)$ decorrelation, the evolution of the excited state equilibrium of the dimer is primarily dictated by the radiative lifetime of the $S_1$ state, the fastest route for return to the ground state. In the conventional analysis of excited state decay, the yield of SF depends on the SF rate constant and the rates of other excited state decay processes: $\Phi_{SF} = k_{SF}/(k_{SF}+k_r+k_{nr})$. For many SF systems (most of which are exothermic), $k_{SF}$ is at least an order of magnitude larger than $k_r$, and the exact value of $k_r$ is largely irrelevant. However, if the radiative and SF rate constants are comparable, as is the case for endothermic SF discovered in 3,10-para connected chromophores, we expect the SF yield will be a sensitive function of $k_r$, and thus strategies to minimize the radiative rate constant, if all other rate constants remain the same, could lead to significant SF yield enhancement.

In molecular dimers $k_r$ is determined by two factors: 1) the intrinsic single chromophore radiative decay (i.e., if removed from the other chromophore of the dimer) and 2) the perturbation that results from the coupling of the chromophores within the dimer. The first is dictated by the strength of the transition dipole moment of the radiative state, assumed to be $S_1$ by Kasha's rule, and can be manipulated through molecular design and substitution. However, for a particular class of chromophores, i.e., acenes with weak transition dipole moments directed along the molecular short axis, $k_r$ falls within a small range of typical values (0.1 – 0.05 ns$^{-1}$). In perylenes, on the other hand, the $S_1$ transition dipole moment is oriented along the long molecular axis, resulting in faster radiative rates (0.2 ns$^{-1}$). For (2), the value can be significantly altered if the $S_1$ transition dipole moments from two or more chromophores interact, resulting in exciton coupling of the excited states, called Davydov splitting in molecular crystals. The magnitude of this interaction is influenced by the relative orientation of the chromophores in the dimer. The two extremes of transition dipole moment coupling are sandwich (H-type) and head-to-tail (J-type), shown in Figure 1. In the latter scenario, the coupling of the transition dipole moments results in augmentation of the total transition dipole moment, and enhancement of the radiative rate.[14]

The phenylalkynyl-perylene oligomers we previously reported exhibit intramolecular head-to-tail coupling of the transition dipole moments. While the SF rate constant increases with the length of the oligomers due to strong electronic coupling through the bridge, the radiative rate constant also gets larger for longer oligomers, thereby competing with SF. Given that the phenylalkynyl-perylene motif is very appealing for photovoltaic applications due to its large extinction coefficient and chemical photostability, we sought to design dimers retaining this structural feature but with improved SF yields through lowering of the radiative rates.

Our molecular design strategy for reducing radiative rates in dimers is twofold: 1) reduction of the intrinsic transition dipole moment of the chromophore by altering the position of alkyne substitution on the perylene, and 2) reduction of the total molecular transition dipole moment by decreasing intramolecular head-to-tail coupling in the dimers. The resultant dimer structures are presented in Figure 1.

Using steady state and time resolved optical spectroscopy we show that the 3,9 substitution pattern on the perylene results in reduced coupling for the dimers compared to the 3,10 pattern, and *ortho* substitution on the benzene linker results in reduced chromophore coupling compared to *para* substitution. Using femtosecond and nanosecond transient absorption (fsTA and nsTA, respectively) we unravel trends in both TT and $T_1$ formation via endothermic SF in the four perylene dimers and provide insights for further direction of molecular design.

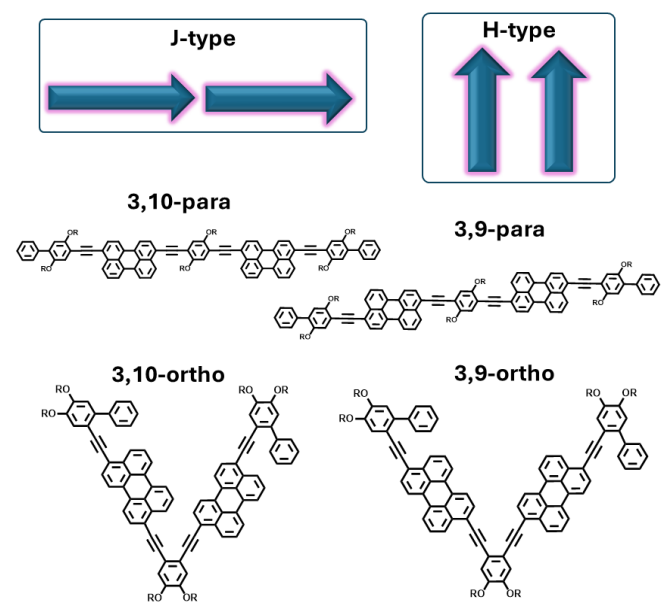

**Figure 1.** Transition dipole alignment basis scenarios (upper) and molecular structures (lower).

## Results and Discussion

The magnitude of chromophore coupling in the dimers can be gleaned from steady state UV-vis and photoluminescence (PL) spectra (Figure 2). The largest red shift compared with the monomer in the steady state spectrum is exhibited by the **3,10-para** dimer, which we reported earlier.[4b] Changing the substitution pattern on the perylene and on the benzene linker results in less pronounced red-shifts, reflecting the extent of chromophore coupling. The least red-shifted spectra are those of **3,9-ortho**, where the coupling is reduced through both the perylene substitution pattern and the geometry of interchromophore interactions.

The shapes of the UV-vis and PL spectra (Figure 2a) shed light on the intramolecular transition dipole moment interactions.

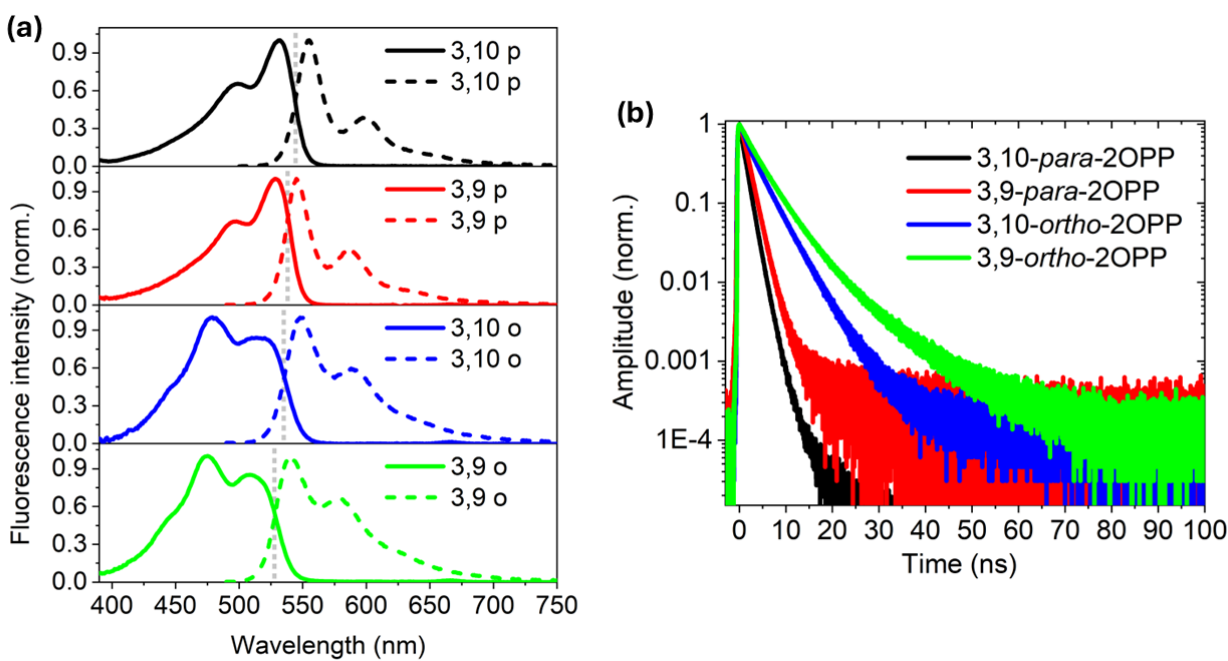

**Figure 2.** UV-vis absorption (solid) and fluorescence (dashed) spectra of the four perylene dimers in THF solution. (b) Time-resolved photoluminescence decay curves for each of the four dimer samples in THF solution.

In para dimers, where transition dipole moments can only couple head to tail, we see an enhancement of the lower energy vibronic feature in the absorption spectrum, and higher energy vibronic peak in the emission spectrum. In the ortho dimers, however, the absorption and emission spectra are not mirror images of each other. While the emission spectra of the ortho dimers exhibit a vibronic progression analogous to the monomers, the absorption spectra show a more prominent second vibronic peak. This type of absorption shape has been observed in other flexible cofacial chromophore dimers and has been assigned to partially H-aggregate-like interactions of the chromophores' transition dipole moments.[15] In the ortho dimers we interpret the clear lack of mirror symmetry to signify that absorption occurs to higher and lower Davydov-split singlet excited states, while the emission occurs only from the lower of the two. The molar extinction spectra (Figure S7) reveal the higher oscillator strength for para connected dimers compared with ortho dimers.

The reduced J-type nature of coupling in the new dimers is indeed reflected in the fluorescence lifetimes and radiative rates (Figure 2b, Table I). The **3,10 para** dimer has the shortest PL lifetime, followed by **3,9 para**, **3,10 ortho**, and **3,9 ortho**. These results suggest that both the substitution position and the transition dipole orientation affect the PL lifetime, but the impact of the relative orientation appears to be greater. The total fluorescence quantum yields (Table 1) reflect the slowing of the radiative rate constants, as they are reduced as lifetimes increase.

**Table 1.** Photophysical properties of the four perylene dimers.

| dimer | $E$ ($S_1$) (eV) | $\Phi_F$ | $\tau_F$ (ns) | $k_r$ (ns$^{-1}$) | $k_{nr}$ (ns$^{-1}$) | $k_{nr}/(k_r+k_{nr})$ |
|---|---|---|---|---|---|---|
| 3,10 p | 2.28 | 0.76 | 1.28 | 0.59 | 0.19 | 0.24 |
| 3,9 p | 2.30 | 0.57 | 1.43 | 0.40 | 0.30 | 0.43 |
| 3,10 o | 2.31 | 0.57 | 2.85 | 0.20 | 0.15 | 0.43 |
| 3,9 o | 2.35 | 0.49 | 3.72 | 0.13 | 0.14 | 0.51 |

In addition to prompt fluorescence, all dimers exhibit some amount of delayed fluorescence, signified by a second slow component in the decay. Delayed fluorescence in chromophore dimers with $E(S_1) \approx 2 \times E(T_1)$ energetics is a reliable sign of intramolecular singlet fission.[16] While kinetic model fitting of PL decays is required for quantifying the yield of SF, a qualitative estimate can be gleaned from comparing the relative amplitudes of prompt and delayed fluorescence decays. The smallest relative amplitude of delayed PL is exhibited by **3,10 para**, followed by **3,10 ortho**, **3,9 para**, and **3,9 ortho**. Fits of the PL decays with a kinetic scheme are found in Figure S1.

To better judge the source of the delayed fluorescence, we performed transient absorption (TA) experiments for THF solutions of these dimers. The normalized femtosecond TA (fsTA) spectra along with the sensitized $T_1$ absorption spectra for all four dimers are shown in Figure 3. In these spectra, negative features in the 500-560 nm range correspond to the ground state bleach (GSB), the negative features in the 575-625 nm range are due to stimulated emission (SE), and the peak at ~800 nm corresponds to the $S_1 \rightarrow S_n$ absorption for all dimers. The triplet absorption spectral shapes depend heavily on the geometry of the dimer. In the para dimers, the $T_1 \rightarrow T_n$ absorption is broad, and ranges from 580 to 650 nm; in the ortho dimers, is more structured from 550 to 700 nm with a peak at 590 and 650 nm for the **3,10** and **3,9** dimers, respectively.

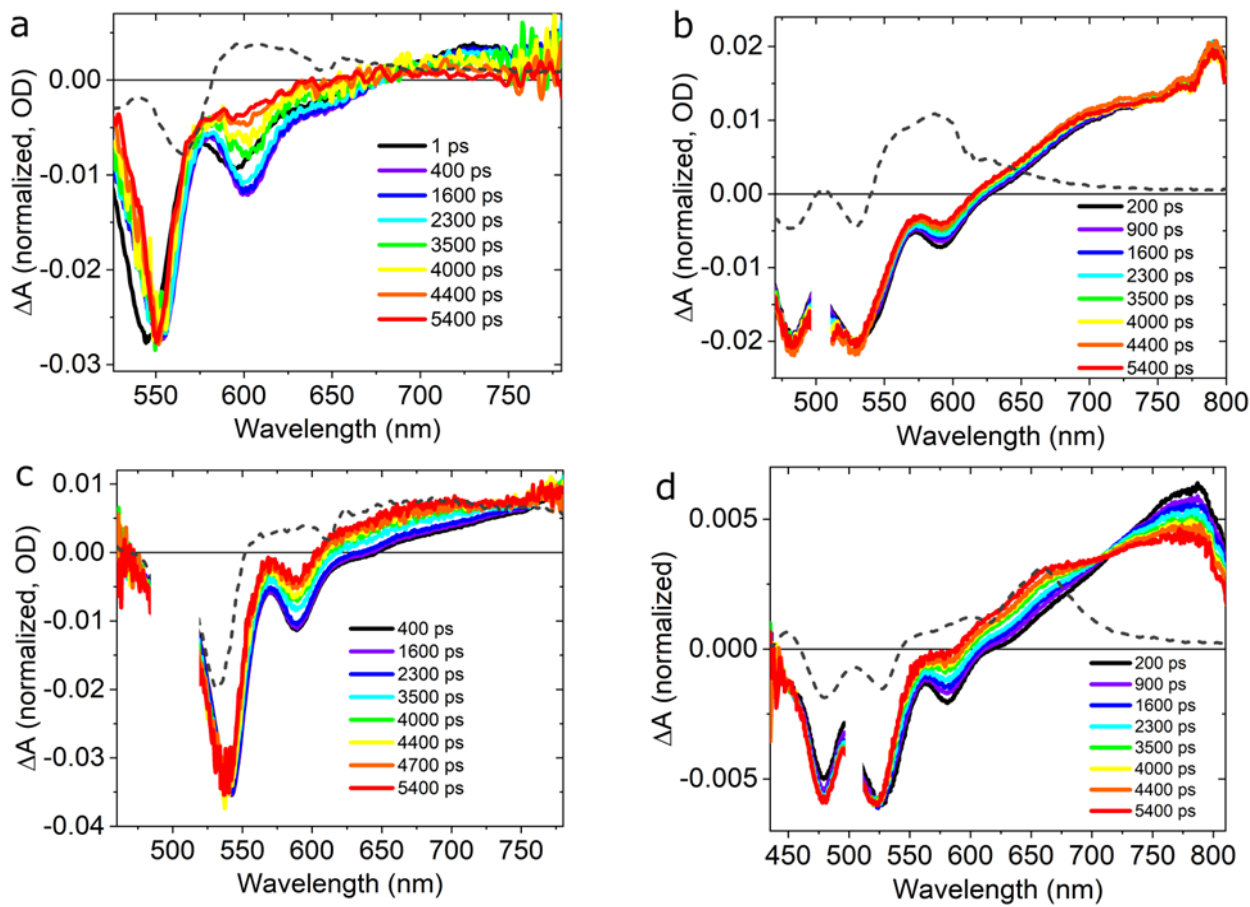

**Figure 3.** fsTA with sensitized triplets for (a) **3,10 para**, (b) **3,10 ortho**, (c) **3,9 para**, and (d) **3,9 ortho.**

The spectral shapes of the excited state absorption for all dimers change to some extent within the 5 ns time window, however, the time dependent spectral change is most pronounced in the **3,9 dimers**. As reported earlier, the **3,10 para** dimer nearly completely returns to the ground state within 5 ns, and the small amplitude of the trailing signal exhibits a GSB, very weak $S_1 \rightarrow S_n$ absorption, and a significantly diminished SE. While no explicit $T_1 \rightarrow T_n$ peak is observed in **3,10 para** dimer absorption, we surmise that the diminished SE signal at later time delays is due to an underlying $T_1 \rightarrow T_n$ absorption. In the **3,10 ortho** dimer, the spectral shape changes slightly at longer time delays, reflecting a weak underlying growth in the region of $T_1 \rightarrow T_n$ absorption, but the late time spectra are still dominated by $S_1 \rightarrow S_n$ absorption along with SE and GSB.

In the **3,9 dimers** the spectral shape change throughout 5 ns is more dramatic than for the **3,10** dimers. The 5 ns trace of the 3,9 para dimer shows a close overlap with the sensitized $T_1 \rightarrow T_n$ absorption; however, unlike the 3,10 para dimer, 3,9 para still has significant amplitude in the region of $S_1 \rightarrow S_n$ absorption at 5 ns. Nonetheless, an isosbestic point is observed at 725 nm between the $T_1$ and $S_1$ absorptive features, indicating a selective conversion between these species that is incomplete on this timescale.

Inspection of TA results up to 5 ns delay provide an initial indication that the reduced radiative rate in ortho dimers results in increased TT formation. Absorptive features that overlap strongly with sensitized $T_1$ are more prominent for ortho dimers; however, TT spectra are not necessarily identical to $T_1$, and significant $S_1$ population remains for the ortho dimers, making it difficult to judge the trends in TT yields accurately.

TA experiments on the ns to μs scale reveal a biphasic triplet decay after the initial $S_1$ decay, including a 1-10 ns component and a decay on the μs timescale (see Figure S4). The longest-lived spectra were compared with long-lived spectra from a triplet sensitization experiment, Figure 4. The comparison reveals a close match between these spectra, indicating a correct assignment to the triplet. Further, the alignment of triplet absorption strength when normalized at the peak of the bleach indicates that just one triplet is present per dimer. This is in contrast to our observations with trimers and tetramers connected via **3,10 para** linkages that showed a twice the triplet strength per unit bleach,[15] indicating triplet pairs vs. single triplets. What is further noteworthy is how distinct the $T_1 \rightarrow T_n$ spectra are, despite being ostensibly due to the same localized triplet species on effectively identical chromophores. It is evident that, even for relatively localized triplets, interchromophore coupling in dimers through either excitonic or wavefunction effects can perturb these spectra significantly from that of the isolated monomer.[17]

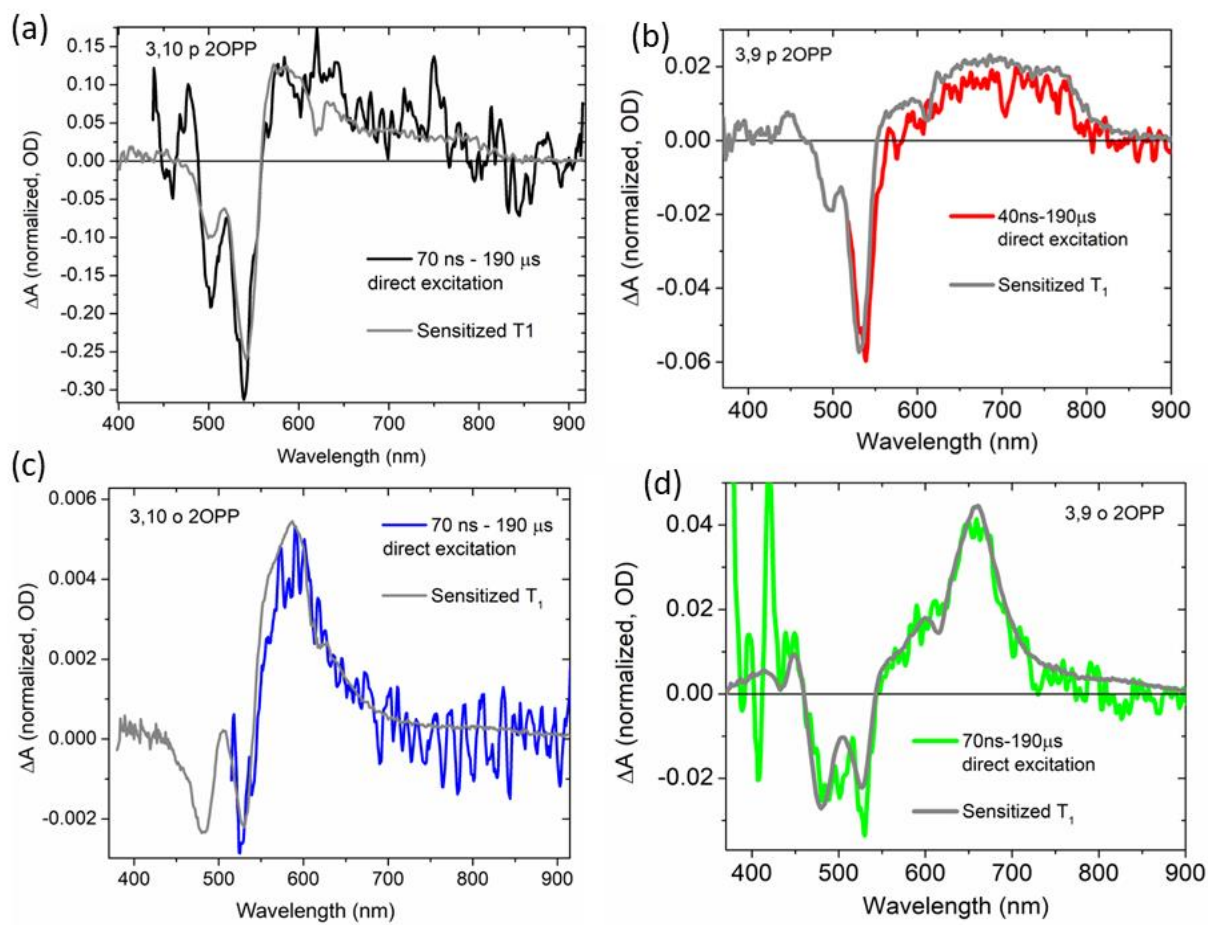

**Figure 4.** nsTA with sensitization and direct excitation for perylene dimers with (a) 3,10 para, (b) 3,9 para, (c) 3,10 ortho, and (d) 3,9 ortho connectivities.

Overall long-lived triplet quantum yields (Table S3) were first estimated for each compound by monitoring the $T_1$ signal strength at long delay times, compared to a standard reference triplet former (See SI for details). $T_1$ formation yields can be quantitatively estimated at long times, well after $S_1$ and TT have decayed. Those yields are surprisingly not maximized for the ortho dimers, suggesting TT→T pathways that also have a dimer geometry dependence. Evidently $S_1$ to $T_1$ via TT involve steps that are not captured by the typical competitive rate constant equation quoted earlier. A trend with estimated endothermicity is also not found, suggesting that an activation barrier to TT formation is not the dominant factor.

We therefore also estimate TT yields at intermediate times by monitoring the associated spectral signature compared with the early time $S_1$ population. These yields are only relative, as extinction coefficient differences have not been taken into account. The results for method depend on a global target fit model, which was performed for the ortho dimers using the 3-component kinetic scheme in Figure S5. The fit parameters are listed in Table S2 and capture the dynamics well for both ortho dimers. The species-associated spectra are shown in Figure S3 and reveal distinct spectral components. The first component matches well with the $S_1$-$S_n$ spectrum (determined from the initial photoexcited state), and the second and third components match $T_1$-$T_n$ determined from sensitized experiments. The second and third components are spectrally similar, with the second tentatively assigned to the putative (TT) species.

Given the fidelity with which the model appears to extract spectra of independent species, the triplet pair yields can be estimated from the population profiles derived from target analysis, shown in Figure S5. It is clear that the **3,9 ortho** dimer achieves the higher (TT) yield than the **3,10 ortho**, with a ratio of roughly 40% to 25%. This can be rationalized based on the radiative decay rate constants, which suggest that (TT) formation is most competitive with fluorescence for the **3,9 ortho** dimer. This analysis of relative yields assumes that the $S_1$ and (TT) extinction coefficients are the same for the 3,9 and 3,10 dimers and cannot reveal the overall $T_1$ yields because those are explicitly assumed to be 100% in the model. $T_1$ quantum yields are determined independently (Table S3) and are reduced compared with (TT) yields because the triplet pair cannot necessarily separate into independent triplets effectively in a system of dimers.

For the **3,9-para** compound, the B and C components are nonzero but nearly identical, suggesting that (TT) is unperturbed compared with $T_1$, similar to linear **3,10 para** trimers and tetramers. As previously described, the excited state scheme for **3,10 para** does not involve separation of (TT) into free triplets, and thus the C component is superfluous, containing only a diminutive residual bleach feature. The nearly zero long-lived triplet yield (Table S3) is consistent with the lack of any evolution beyond the (TT) state.

Based on the prominent triplet features observed for the **3,9 ortho** dimer (and the similar though weaker features for other dimers) we surmise that population in $S_1$ evolves to produce triplet pairs on the ns time scale. The reversibility of the transition to allow a fast $^1(TT)$ to $S_1$ transition enables the formation of an equilibrium between $S_1$ and $^1(TT)$, which can be destroyed through $S_1$ decay to the ground state, or $^1(TT)$ depopulation. The latter process primarily involves triplet migration for oligomers to produce $T_1 + T_1$, but in dimers where this route is unavailable, decoherence could result from a transition to the $^3(TT)$ state. In aligned chromophore systems, this pathway should be disallowed,[18] but available non-parallel conformations would allow $^3(TT)$ formation via the $^5(TT)$ state, albeit on a relatively slow timescale. The eventual formation of one $T_1$ per dimer suggests that this pathway is likely, as subsequent $^3(TT)$ to $T_1$ formation would be spin-allowed, facile, and consistent with our overall observations. The competition between access to $^3TT$ and a $TT \rightarrow S_1$ or $TT \rightarrow S_0$ pathway is likely the determining factor in eventual $T_1$ yields. Although the dimer juxtaposition may dictate the branching of these pathways, because the yields are fated to be far less than 100%, further analysis should proceed on systems with multiple chromophores where triplets can truly separate.

## Conclusion

Endothermic SF is affected by the singlet fission energy balance and the lifetime of the excited state in a series of dimers with varying connectivity. Although systematically varying radiative rate constant does produce a trend in the dominance of TT population, elaborating these findings to trimers or tetramers with geometries other than the 3,10 para connectivity is a future goal that is likely to succeed in producing significantly higher free triplet yields in perylene systems than previously reported. Further, loss pathways associated with access to $^3TT$ (and is associated internal conversion loss pathway to $T_1$) could be examined with time-resolved EPR or related magnetic resonance techniques. Reducing these pathways could lengthen TT lifetimes for eventual harvesting by TT dissociation or direct TT transfer to an acceptor.[19]

## Experimental Section

**Optical Spectroscopy**. The UV visible spectra were recorded on a Varian Cary 50 Conc UV Vis spectrophotometer. The steady-state emission at room temperature was measured with a Horiba Jobin Yvon FluoroMax 4 spectrofluorometer. Fluorescence quantum yields were determined by using a solution of rhodamine 6G in ethanol (O.D. <0.1) as a standard with PLQY = 0.95.
Transient absorption datasets were acquired using a Coherent Libra Ti:Sapphire laser, with an output of 800 nm at 1 kHz. A TOPAS-C OPA was used to generate the ~150 fs pump pulse tuned from 500 to 650 nm for these studies to excite the peak of the sample absorption. The pump pulse energy was typically 5 120 nJ, and the pump spot size was found to be approximately 300 μm obtained by measuring the power before and after a series of pinholes at the sample position and fitting assuming a Gaussian spatial distribution. In an Ultrafast Systems Helios Spectrometer, a small amount of 800 nm light was used to pump a 1 mm sapphire crystal to generate 450–800 nm probe light for UV-VIS TA. A delay up to 5 ns can be achieved with the Helios. For delays up to 400 μs, an Ultrafast Systems EOS was used, which employs an electronically delayed probe continuum pulse and the same excitation characteristics as the Helios. A Janis liquid nitrogen cryostat with a customized sample holder was used for 77 K measurements.

TRPL measurements were made using a supercontinuum fiber laser (Fianium, SC-450-PP) operating at 1 MHz as the excitation source. The excitation wavelength was chosen using an acousto-optic tunable filter to be 500 nm with of pulse energy of approximately 0.1 nJ. A streak camera for detection in the range of 500-700 nm (Hamamatsu C10910-04) was used to detect time-resolved spectra. The instrument response function for a time range of 0-20 ns is approximately 200 ps. The region near the peak sample emission was integrated to produce the fluorescence decay curves.

Solution samples were prepared in the glovebox under an atmosphere of $N_2$. The compounds were dissolved in the desired solvent and passed through a 0.22 μm PTFE syringe filter. Fluorescence experiments were done in 1 cm cuvettes with O.D. of <0.1. Transient absorption samples were made in 2 mm cuvettes with sample O.D. in the 0.2 0.6 range.

## Supporting Information

Synthesis information, full transient absorption data sets, global fitting information, triplet yield determination

## Acknowledgements

This work was supported by the Solar Photochemistry Program of the U.S. Department of Energy, Office of Basic Energy Sciences, Division of Chemical Sciences, Biosciences, and Geosciences. KL acknowledges funding for precursor synthesis from the SCGSR program of the United States Department of Energy, Office of Basic Energy Sciences. JK and SO acknowledge the SULI program of the DOE. This work was authored by Alliance for Sustainable Energy, Limited Liability Company, the manager and operator of the National Renewable Energy Laboratory under Contract No. DE‐AC36‐08GO28308. The views expressed in the article do not necessarily represent the views of the Department of Energy or the U.S. Government. The U.S. Government retains and the publisher, by accepting the article for publication, acknowledges that the U.S. Government retains a nonexclusive, paid‐up, irrevocable, worldwide license to publish or reproduce the published form of this work, or allow others to do so, for U.S. Government purposes.